# Sub one percent mass fractions of young stars in red massive galaxies

Núria Salvador-Rusiñol[1,2*], Alexandre Vazdekis[1,2], Francesco La Barbera[3], Michael A. Beasley[1,2], Ignacio Ferreras[1,2,4], Andrea Negri[1,2], Claudio Dalla Vecchia[1,2]

[1]Instituto de Astrofísica de Canarias, E-38200 La Laguna, Tenerife, Spain.

[2]Departamento de Astrofísica, Universidad de La Laguna, E-38205 La Laguna, Tenerife, Spain.

[3]INAF – Osservatorio Astronomico di Capodimonte, I-80131 Napoli, Italy.

[4]MSSL, University College London, Holmbury St Mary, Dorking, Surrey RH5 6NT, UK.

*nsalva@iac.es

**Early-type galaxies are considered to be the end-products of massive galaxy formation[1]. Optical spectroscopic studies reveal that massive early-type galaxies formed the bulk of their stars over short timescales (≤1 Gyr) and at high redshift (z≤2), followed by passive evolution to the present[2]. However, their optical spectra are insensitive to constrain small episodes of recent star formation, since they are dominated by old stars. Fortunately, this problem can be tackled in the ultraviolet range. While recent studies that make use of ultraviolet absorption lines have suggested the presence of young stars in a few early-type galaxies[3], the age and mass fractions of young stars and their dependence on galaxy mass, is unknown. Here we report a detailed study of these young stellar populations, from high-quality stacked spectra of 28,663 galaxies from the BOSS survey[4], analysing optical and ultraviolet absorption lines simultaneously. We find that residual star formation is ubiquitous in massive early-type galaxies, measuring average mass fractions of ~0.5% in young stars in the last 2 Gyr of their evolution. This fraction shows a decreasing trend with galaxy stellar mass, consistent with a down-sizing scenario[5]. We also find that synthetic galaxies from stateof-the-art cosmological numerical simulations[6] significantly overproduce both intermediate and young stellar populations. Therefore, our results pose stringent constraints on numerical simulations of galaxy formation[6,7].**

Observations of early–type galaxies (ETGs) with GALEX and Sloan Digital Sky Survey (SDSS) photometry show a large scatter in their ultraviolet (UV) and optical colours and magnitudes, possibly indicative of some residual star formation in the last Gyr[8]. We used data from the SDSS–III Baryon Oscillation Spectroscopic Survey[4] (BOSS) comprising spectra of luminous red galaxies. These are passive systems, typically featuring a bulge-dominated morphology and are mostly ETGs[9]. We selected galaxies in the redshift range 0.35≤z≤0.6, so that the spectra map rest-frame wavelengths between 2,500 and 6,500 Å, providing both optical and near-UV (NUV) data. Further, we chose massive galaxies with stellar velocity dispersion (σ) between 220 and 340 kms$^{-1}$. We adopted the colour cut (g-i)>2.35 to remove late-type galaxy contamination[9]. The final sample comprises 28,663 galaxies (Extended Data Figure 1) with median redshift z~0.38. Due to the BOSS fiber size, at this redshift we sample the central regions of these galaxies (~10 kpc). The use of a large sample allows us, for the first time, to average over individual variations and provides a clear picture of the incidence and mass fractions of very young stellar populations in massive ETGs.

The signal to noise ratios (SNR) of the individual BOSS spectra are insufficient for detailed stellar population analysis so we stacked them by σ into 7 bins (Extended Data Figure 2). Stellar velocity dispersion correlates strongly with population properties of ETGs[10,11] and relates directly to their dynamical stellar masses[12,13]. From the optical Hβ$_o$ and [MgFe]' line-strength indices, we derive old mean luminosity-weighted stellar ages (MLWA) and high metallicities, suggesting that the stellar component of massive galaxies were formed at early epochs[2], and become older and more metal-rich with increasing mass.

To analyse the BOSS stacks, we employ the E–MILES[3] single stellar population (SSP) models (see Methods). We measure 14 absorption line indices covering the NUV and optical ranges, which are defined in Supplementary Table 1. Figure 1 shows some representative indices (see Extended Data Figure 4 for the remaining set). Note that while the observed optical indices are in excellent agreement with the old SSP model predictions, there is an increasing difference between model predictions and observed data towards bluer indices. This behaviour provides clear evidence of the inability of a pure old SSP to fully describe the star formation history (SFH) of massive ETGs and calls for an additional component[3,14]. The decrease of the NUV indices cannot be due to intermediate-age stellar populations, because they barely contribute in the NUV. In order to understand this additional component, we adopt two simple parametrizations of the SFH of a massive and old ETG as a superposition of an old and a young component. The old component is described by a SSP. The young component – tracing the recent star formation episodes – is given either by an additional, SSP (hereafter, called 2SSPs model), or by a composite population following a constant star formation rate (SFR) within the last 2 Gyr of galaxy evolution (hereafter, called 1SSP+cSFR model), which is more representative of the real SFH of a massive ETG than the 2SSPs model. The 2SSPs serves to understand the origin of the departure of the UV absorption index strengths in relation to the 1SSP that describes well the line indices in the optical. This model is a simplistic assumption, as it assumes the young component forming instantaneously at a given specific time, while we actually expect any residual star-formation happening over an extended time interval. For this reason, we believe that the 1SSP+cSFR assumption for the young component is a more realistic approximation. These two modelling approaches are compared with the stacked spectra by fitting 14 absorption line indices to estimate the young stellar contribution in massive ETGs.

The 2SSPs model assumes that the spectrum of a massive ETG comprises two individual SSPs, with appropriate weights (see Methods). The best-fitting model, shown as red stars in Figure 1, reveals a small but detectable presence of young stars (see Extended Data Figure 5). Extended Data Figure 3 shows the probability distribution functions (PDFs) between each pair of fitted parameters for a given stack. Panels a) and b) of Figure 2 show the derived young mass fractions and their MLWA as a function of $\sigma$, suggesting that massive ETGs have small mass fractions of young components, between 0.09% and 0.15%, decreasing with stellar mass. The average young age also shows a trend with velocity dispersion, from 109 Myr at the lowest $\sigma$ stack to 65 Myr in the most massive galaxies (Supplementary Table 2). Blue shaded areas in the figure show how results change when giving different relative weights to the optical and NUV indices in the fitting process, varying them from 0% to 100%. This allows for the exploration of extreme cases when one has more information from one spectral range or the other. In most cases, the largest differences with respect to the reference solution in the derived values is obtained when giving more weight to the NUV indices, due to their age degeneracy. We note that although the effect of internal dust reddening is expected to be more important in lower mass galaxies, its role in the analysis of the narrow NUV spectral indices studied here is negligible (see Methods).

Going further, we confront the observed index-strengths with the predictions of the 1SSP+cSFR approach. We find the total mass fraction of stars formed over the last 2 Gyr is 0.48% in the most massive galaxy stack (i.e., highest $\sigma$) with a weak increase to 0.58% at the lowest $\sigma$. From this we determine an average specific SFR (sSFR) of ~2.6x~$10^{-12} yr^{-1}$. The 1SSP+cSFR best-fitting indices are shown as blue open triangles in Figure 1 and Extended Data Figure 4.

Thanks to the unprecedented sensitivity of our analysis due to the high SNR spectra and a large sample size, which allows us to cover a wider coverage of masses with respect to previous works, our results show, for the first time, that the population of massive ETGs at redshift z~0.4 hosts a young stars component, at the sub-one percent level, whose contribution anti-correlates with galaxy stellar mass. This trend can be interpreted within a down-sizing scenario[5], which argues that the formation of stars in less massive systems extends over longer time-scales with respect to more massive galaxies[15]. The fact that in the lower mass stacks we obtain older ages for the young component in the 2SSPs model might be indicative of a more prominent intermediate-age stellar population, with respect to more massive galaxies, i.e., a longer formation timescale. These results impose fundamental constraints on the later stages of the evolution of massive ETGs.

At different evolution epochs, LEGA-C[16] inferred that the equivalent ages for the oldest galaxies with stellar mass > $10^{11}$ $M_\odot$, are ~5.5 Gyr old at z~0.8[17], which matches reasonably well the equivalent age of our most massive bin at z~0.4, taking into account the difference between the age of the Universe at both epochs (3 Gyr). In the nearby Universe, ATLAS-3D[18] reveal these galaxies as being fundamentally old (~13 Gyr) and metal-rich and have an average sSFR of ~$10^{-12}$ $yr^{-1}$ at z~0.4[19], in good agreement with our estimations from the 1SSP+cSFR approach.

However, the NUV is not only sensitive to young stars but also to hot, old stars such as from the Post-Asymptotic Giant Branch (PAGB) evolutionary stage[14]. As a check, we performed an analysis combining an old SSP with the spectrum of a high temperature (10,000 – 25,000 K) star, which resembles a PAGB star (see Methods). We find a ~4% light contribution from the PAGB in the optical (V-band) would be required to fit the data. The contribution of the PAGB according to SSP models[20,21] is less than 1x$10^{-6}$, in mass, which gives significantly less than 1% in light in the V–band. This implies that the contribution from hot stars is likely due to a young stellar component, rather than PAGB stars. However, it might be the case that both young and PAGB contributions are at work, in which case our young mass fraction estimates should be considered, to some extent, as an upper limit. Additionally, we expect a negligible contamination from the interstellar medium to the NUV absorption features. For reference, the Fe3000 index is unaffected by the interstellar absorption[22], while this index shows a deviation with respect to an old component, which is well fitted with an additional small young component.

At present, the physical processes involved in the residual star formation activity found in massive ETGs are not well understood. It is known that feedback mechanisms must be invoked to prevent protracted star formation in massive galaxies. Such extended SFHs are expected from a naive connection between dark matter growth and stellar mass growth in a hierarchical context. Feedback will alter this connection, by preventing the infalling gas from cooling and forming stars. Physical mechanisms such as AGN activity and type Ia supernovae are known to influence the SFR in massive galaxies[23]. Supernova feedback is not thought capable of driving this phenomenon by itself to produce quiescent galaxies[24]. AGN feedback provides a plausible solution, as proposed by numerical simulations that suggest it is the most likely process by which the available cold gas is heated and expelled by their energetic jets in massive ETGs[25,26]. These processes are more effective in the most massive systems, agreeing well with our results, where the fraction in young stars is anti-correlated with the stellar mass of the galaxy. However, our findings also show that quenching mechanisms are not able to completely stop star formation in the central regions of ETGs, even in the most massive galaxies, at the observed stringent levels. Therefore, additional processes must be present to control the cold gas supply that results in the formation of new stars at late times in massive ETGs. Cold gas has been found in atomic and molecular form in various samples of ETGs[27], a result that has been connected to the SFH over the last few Gyr[28].

Our sample comprises ETGs with a mean stellar mass higher than $\sim 10^{11}$ $M_\odot$[29], (Supplementary Table 2). Such massive galaxies mostly live in groups. Half of our sample is expected to inhabit galaxy clusters (see Methods). Therefore, the residual star formation found may be caused not only by intrinsic processes, such as gas returned to the interstellar medium from late phases of stellar evolution[30], but also might be triggered by ex-situ contributions that reach central galaxy regions.

Numerical simulations of galaxy formation help to understand the role of the complex physics at play. We use the latest, state-of-the-art high-resolution cosmological hydrodynamic modelling from the EAGLE simulation[6] to explore our findings. The EAGLE simulation targets average density environments, typical of those inhabited by BOSS galaxies. To create a simulated sample whose properties should be compatible with those of our stacked spectra, we applied the BOSS colour criteria and selected EAGLE galaxies within our σ range. From the set of 3,186 EAGLE galaxies with stellar mass above $10^{10}$ $M_\odot$, only 26 galaxies satisfy those criteria. Their optical H$\beta_o$ and [MgFe]' indices suggest substantially younger MLWAs than those derived from the stacks (Figure 3). For this reason, with the exception of the two objects that match the data (which also happen to be the oldest), simulated galaxies are excluded when applying the colour cut (g-i)>2.35 due to their young ages. Figure 3 also illustrates the SFHs of a young and old simulated galaxy in panels a) and b), respectively. Clearly, the younger galaxy has a more extended stellar age distribution and contains more intermediate stellar populations than the older galaxy and, as a consequence, H$\beta_o$ is stronger in the younger galaxy. Observations indicate that massive ETGs formed the bulk of their stellar populations at earlier epochs without a significant component of intermediate and young age stars, in sharp contrast with the simulations, which overproduce recent star formation. Hence, the SFHs of our stacked spectra should broadly resemble the one shown for the old simulated galaxy. The mass fraction in young stars formed over the last 2 Gyr from the SFHs of the two EAGLE galaxies that best compare with the observations are 9.23% and 0.07%, from left to right in the grid plot.

An important source of uncertainty in the present work is the lack of simulated galaxies with SFHs resembling the observations. Only one simulated galaxy appears to approach the behaviour of the BOSS galaxies, in terms of young mass fractions. This result urgently needs to be better understood with a larger sample of simulated massive galaxies. Therefore, our results suggest a revision of how numerical models control feedback and the baryonic processes of galaxy formation in massive BOSS-type galaxies. The observations presented here provide the most stringent limits on recent star formation in ETGs to date and represent a key benchmark that theoretical models must satisfy in order to provide a more realistic description of galaxy formation and evolution.

**Methods**

**Data selection.** We used galaxy spectra from the SDSS-III/DR12 BOSS survey[4]. This survey mapped luminous red galaxies, representing predominantly massive quiescent galaxies with an early-type morphology[9]. We restricted the sample between redshift 0.35<z<0.6, with SNR > 7 in the SDSS-r band (Extended Data Figure 1), in order to select objects that cover the targeted NUV and optical line strengths and to avoid excessively noisy spectra. We also restricted our selection in stellar velocity dispersion, σ>220 kms$^{-1}$, to select massive galaxies. At such velocity dispersion, the sample should be practically free from late-type galaxies[31]. We additionally adopt the colour cut (g-i)>2.35 to remove remaining late-type galaxy contamination[9]. The stellar velocity dispersion of the individual galaxies was obtained with pPXF[32] in the wavelength range 4,500 – 6,500 Å, covering a number of absorption features suitable to measure the kinematics[33].

The BOSS sample that satisfies the above criteria comprises 28,663 galaxies and primarily consists of ETGs. No environmental criteria have been taken into account in the selection process, therefore our sample includes not only field galaxies but also galaxies potentially in galaxy clusters (see below). The diameter of the BOSS fibers maps into 2 arcsec on the sky, covering ~10 kpc aperture at the median redshift distance of the sample (z~0.38), so we are looking at the central regions of the galaxies. Note that the distribution is biased towards lower redshifts. We adopt a flat λ-CDM cosmology, with a Hubble constant $H_0$=70 kms$^{-1}$Mpc$^{-1}$ and $\Omega_m$= 0.30.

The BOSS spectra are given in vacuum wavelengths. We convert them to the air system using the IAU standard equation[34]. We also applied a Milky Way dust extinction correction following the standard Cardelli law[35] with $R_V = 3.1$ and the official SDSS foreground dust extinction values (g–band), $A_g$, transformed into a colour excess via[36] $E(B-V) = A_g/3.793$. Note that the effect of dust is virtually negligible in our study since we analysed a set of spectral features which encompass a narrow spectral region with respect to the variations in the extinction law (see below).

**Stacking the spectra.** Individual spectra have too low SNR to derive reliable information from some of the line strengths, especially in the NUV (SNR~2 Å$^{-1}$). Our detailed stellar population analysis requires at least a SNR of 15 Å$^{-1}$ for measuring some NUV indices to disentangle tiny young stellar mass fractions contributions. Therefore, stacking is the best approach to increase SNR, given that our sample comprises a large number of spectra. We create the stacks by bringing the data to rest-frame wavelengths and co-adding all the spectra that correspond to the same group using a grid with fixed sampling in log-wavelength, $\Delta \log(\lambda/Å) = 10^{-4}$. The flux per pixel is shared between neighbouring reference pixels, proportional to their level of overlap with the original, de-redshifted pixel. Given the redshift range of the sample, the shifted data spans the wavelength range 2,500 - 6,500 Å. The data are grouped into seven stacks, binned in velocity dispersion σ from 220 to 340 kms$^{-1}$ in steps $\Delta\sigma$ =10 kms$^{-1}$ for the first 4 stacks, and slightly wider for the stacks corresponding to the three highest velocity dispersions: $\Delta\sigma$ =20 kms$^{-1}$ and $\Delta\sigma$ 40 kms$^{-1}$, to achieve a roughly constant number of individual spectra per bin and sufficient SNR to perform the analysis. The error on the flux is coadded (in quadrature) in the same way as the flux, from the individual values at each wavelength. The stacks achieve very high SNR, above 30 in the NUV and several hundred in the optical (Extended Data Figure 2). Most importantly, the stacking procedure allows us to obtain results that apply to the general ETG population, by washing out galaxy-to-galaxy variations at fixed velocity dispersion. General properties of each σ bin are listed in Supplementary Table 2.

**Line-strength absorption features and stellar population models.** The high SNR of our stacked spectra allows us to study a set of 14 targeted absorption indices (defined in Supplementary Table 1), covering a wide spectral range, from the NUV to the optical. These indices include metallicity indicators and Balmer line age-sensitive indicators in the optical range, and NUV indices sensitive to the young stellar components. Previous studies used optical line indices to derive stellar population and kinematic properties of ETGs[37,38]. A key novelty of the analysis presented here is the simultaneous fitting of optical and NUV spectral indices. Our fitting procedure excludes those features that are highly sensitive to the [Mg/Fe] abundance ratio, such as Mg2800 and Fe3619, as massive ETGs are enhanced in magnesium relative to iron[39]. In addition, Mg2800 is potentially affected by chromospheric emission from the stellar atmospheres. To constrain the stellar content, we employed the state-of-the-art stellar population models E-MILES[3] – covering a wide range of stellar population parameters at moderately high spectral resolution – to perform detailed fits of the set of line indices and derive equivalent ages and metallicities of the stellar populations. E–MILES is currently the only set of public available models entirely based on empirical stellar libraries covering the UV range. We used the version of E–MILES that extends down to 6.3 Myr[40] and employ the Padova isochrones[41]. The grid of models spans metallicities, $-2.32 \leq [M/H] \leq +0.22$, with ages ranging from 6.3 Myr to 14 Gyr. We adopt a low-mass tapered bimodal initial mass function with logarithmic slope of 1.3, and scaled-solar abundances[42]. We performed Gaussian smoothing in both SSP models and stacked spectra at a common central velocity dispersion of σ = 340 kms$^{-1}$ to remove the effect of variations of the effective spectral resolution on the line strengths.

Figure 1 highlights the advantage of combining indices from both spectral windows. The NUV features are, however, age-degenerate (note the shallow slope of the solid and dashed lines in the NUV indices shown in the panels a) and b) of Figure 1), while a combination of optical and NUV features effectively breaks this degeneracy, allowing us to discriminate between an SSP and a more complex SFH, as the UV is extremely biased towards the minute fractions of young stars present.

**Estimating the errors of the spectral indices.** Two sources of uncertainty are considered in the measured indices. First, the Poissonian noise, which is very small as a result of the high SNR of the stacked spectra. Second, the fluctuations in a given index expected from variations among the galaxies in a given stack. This variation is evaluated via bootstrapping[43], by performing 1,000 realizations of the stacked spectra across the full galaxy sample, selecting random spectra from the original sample (with replacement) for each velocity dispersion bin. The measured indices in the bootstrap form a sampling distribution for each index, from which we define the sampling uncertainty as the mean of the difference between the median of the subset and the 5$^{th}$ and 95$^{th}$ percentiles. We find that the intrinsic variation within the index values of each realization is relatively small, with an average standard deviation of ~0.07 with respect to the median value. Finally, the uncertainty due to the bootstrap resampling is added in quadrature to the Poissonian noise, giving the total estimated error on the line-strength indices used in this work.

**A first look at the underlying stellar populations.** As a first approximation, we constrain the global properties of the stellar populations of each stacked spectrum by fitting with SSP models using two widely used optical indicators, Hβ$_o$ and [MgFe]'. These serve as proxies for the MLWA and metallicities of the stellar population, respectively. Note that we take a single SSP as an approximation for the whole stellar population of the ETGs, representing the global contribution to the spectra. As a visual guide, Figure 3 shows the grid of stellar population model predictions for these two indices for a range of ages and metallicities. The observed strengths indicate an overall old and metal-rich ([M/H] > 0) population. The inferred stellar population parameters are listed in Supplementary Table 2. The age of the stellar populations of ETGs correlate with the stellar velocity dispersion whereby more massive (i.e., higher velocity dispersion) galaxies are older. Moreover, we note that the age obtained for the most massive stack (the oldest) is 8.6±0.5 Gyr, compatible with the age of the Universe at the median redshift of the sample (z~0.38$), namely 9.4 Gyr[44].

ETG spectra may include nebular emission, revealing the presence of ionised gas. Although weaker than in late-type galaxies, we are required to correct for this component, especially those features based on the hydrogen recombination lines – in particular $H\beta_o$ – as even weak emission may partly fill the absorption feature, affecting the estimated age. To correct the $H\beta_o$ line-strength for nebular emission, we subtract a best-fit SSP model spectrum in the region of $H\beta_o$ from each stack and measure the residual of the index. We typically find a small but detectable nebular emission ($\Delta H\beta_o \sim$ 0.07 Å, on average), lower in the most massive bins, but significant enough to change the estimated ages of their stellar populations – at most by +0.8 Gyr at the lowest velocity dispersion range. This residual is added to the observed index value to derive the age. We do not correct the higher order Balmer line strengths in our analysis ($H_\gamma$ and $H_\delta$) as the contribution from nebular emission – directly measured from the residuals – is negligible.

In addition, we have tested that the measured young mass fractions we find are not biased by a small fraction of ETGs that may have recently experienced some star formation event. For this purpose, we measured the $H\beta_o$ and the [OIII] lines in the individual BOSS spectra to find that a very small fraction of galaxies (only ~ 0.5% on average) has significant emission in those lines. To guarantee that these galaxies are not biasing the results, we repeat the fitting process by excluding these targets from the stacking procedure. The young mass fractions turned out to be the same as those obtained for the whole sample. Therefore, we conclude that the young stellar component mass fraction is typical for the whole population of ETGs, confirming the robustness of our results.

**Impact of dust on the stacked spectra.** The spectral continuum in the NUV region can be significantly affected by dust intrinsic to the galaxies. Note the foreground attenuation from the Milky Way is already corrected for in the individual spectra. The wavelength-dependent dust attenuation law is modelled by a smooth function along with a few broad resonant features. Therefore, dust will mostly affect measurements covering large wavelength intervals, e.g. broadband photometry. Narrow spectral features will be, at most, weakly affected, especially in massive ETGs, where the dust content is minimal. Moreover, all our indices define the local pseudo-continuum with a blue and a red sideband, further decreasing the potential impact of dust. This is an important advantage of this analysis over approaches based on broad-band colours. We check the effect of dust on our measurements by artificially reddening the stacked spectrum corresponding to the lowest velocity dispersion range – lower mass galaxies are expected to have the highest dust content – adopting an extinction curve with $R_v$ = 3.0, typical of ETGs[45] and $A_v$ = 0.5$ (note that this is a large extinction for ETGs). The indices obtained in this dusty stack do not differ significantly from the original ones. The largest variations are found in the BL2720 and BL2740 features, which increase by ~ 12% and 3%, respectively, but the other NUV indices change less than 1%. The optical indices are completely unaffected by dust.

**SFH parametrization of ETGs.** While the age-sensitive index $H\beta_o$ shows old ages at z~0.4, consistent with previous works[15,19,46], there is a well-established relationship between galaxy mass and its SFH. More massive galaxies exhibit narrower SFHs that peak in their star formation at older ages than their less massive counterparts. In this case, a pure old SSP is a good first approximation for a galaxy whose mass is larger than $10^{11}$ $M_\odot$, such as the stacked spectra analysed here. However, as shown in Figure 1 and Extended Data Figure 4, the NUV data clearly indicates the inability of an old SSP to fully describe the SFH of massive ETGs, and calls for an additional component[3,14] to be fitted. This component needs to be a very hot stellar population responsible for the decrease of NUV absorption features possibly due to young stars. Here we aim to quantify the upper limit mass fraction of very young stellar populations in massive ETGs at z~0.4. For this purpose, the SFH has been approached by parameterising it in two different simple ways: a young SSP (2SSPs model) and a constant SFR in the last 2 Gyr (1SSP+cSFR model), both on top of an old SSP, explained in more detail in the following sections. The deviation of NUV features from the old SSP model predictions cannot be caused by intermediate-age stellar populations because they barely contribute in the NUV. If such contributions were significant they would alter the optical indices leading to younger MLWA measured with the $H\beta_o$ vs. [MgFe]' diagram (as happens to most of the EAGLE galaxies shown in Figure 3). Thus, intermediate stellar populations are not parametrized in our different SFH choices, as our aim is to constrain the relative fraction of the young stellar contribution.

We explore a parameterization of the SFH adopting an exponentially declining SFH, SFR(t) = exp(–t/ τ) by varying the declining time τ from 0.1 and 2 Gyr. Note that such a modelling approach does not account for any recycled material from stellar evolution and these fractions would change if the recycled material was included. However, the 2SSPs and 1SSP+cSFR approaches only parametrize the mass fractions of some periods of time of the SFH of a galaxy, and whether these stars come from pristine or recycled material cannot be derived from this work. The SFH model templates are generated with E–MILES models assuming a Kroupa initial mass function and solar metallicity. Comparison between data and models indicate a best-fit model with τ =1.2, giving 0.57% total stellar mass formed in the last 2 Gyr. This implies sSFR of 2.9-12 $yr^{-1}$ which matches that derived from the 1SSP+cSFR model. Hence, our results regarding the mass fraction contribution of youngest populations are very robust with respect to the assumed parametrization of the ETGs' SFH.

**Fitting methodology.** For each stacked spectrum, we employ a Markov Chain Monte Carlo (MCMC) algorithm to simultaneously fit the set of 14 optical and NUV line indices with the predictions from the two different modelling approaches described below. The best-fitting solution is obtained by maximizing the log-likelihood function $\ln \mathcal{L}$ given the observed set of indices (I), which is equal to the probability P for model parameters ($\theta$) of getting the observed indices I:

$$P(I|\theta) = \ln \mathcal{L}(\theta|I)$$

where the likelihood is defined by $\ln \mathcal{L}(\theta|I) \propto -\frac{\chi^2}{2}$, where the $\chi^2$ statistic is:

$$\chi^2 = \sum_{i=1}^{N} \left( \frac{I_{model_i} - I_{stack_i}}{\sigma_{stack_i}} \right)^2$$

where the subscript $i$ refers to the $i^{th}$ line index, $I_{model_i}$ and $I_{stack_i}$ are the model and observed indices, respectively, and $\sigma_{stack_i}$ represents the index uncertainty.

We compute the sampling with the *emcee* package[47], a Python implementation of an ensemble sampler[48]. We perform 10,000 model realizations sampling the PDF through 100 walkers exploring the parameter space. 1,000 burn-in steps are used. We use the marginalized PDFs for the estimation of the best-fitting parameters, i.e., taking into account all the probabilities that we get for a given parameter through the other parameters, and assuming the median as the best-fit value and the 16$^{th}$ and 84$^{th}$ percentiles as the uncertainties. Extended Data Figure 3 shows an example of the marginalized PDFs as histograms, and the two-dimensional contours represent the joint PDFs between a given pair of parameters.

Additionally, we also explore how the best-fitting results change when giving different relative weights to the optical and NUV indices, separating the $\chi^2$ statistic at 3,400 Å, slightly rescaling the error bars so that each set of indices has a given weight on $\chi^2$. We varied the weights from 100% to 0% in the $\chi^2$ minimization within the index uncertainties at the 1–$\sigma$ level. The variation is shown by the shaded region in the Figure 2. Varying these weights give a wider range of solutions, and we regard this as a conservative approach.

**Fitting approach 1: 2SSPs modelling.** We fit the data with a 4-parameter model that consists of a superposition of two simple stellar populations. The parameters control the age of the old component (between 1 and 9.4 Gyr; the latter is the age of the Universe at the median redshift of the sample), the age of the young component (between 6.3 Myr and 1 Gyr; the former is the minimum age of the E-MILES SSP models), the metallicity of the old component (between -0.2 and +0.22) and the relative contribution of the young component, measured by mass. The metallicity of the young component is fixed at solar since a small change in metallicity can be compensated by a much smaller change in the age of the young stellar population being difficult to estimate it. In addition, the hottest stars, particularly those around the turn-off with effective temperature $T_{eff} > 10,000$ K, which have little dependence on the metallicity, are the major contributors to the NUV.

This approach is motivated by the fact that massive ETGs formed the vast majority of their stars at early times, over a relatively short timescale ($\lesssim 1 - 2$ Gyr, consistent with their high [Mg/Fe] and old MLWA). Consequently, the stellar populations originated after the main episode of formation do not contribute significantly to the flux in the optical range, as their relative mass fraction is small and their corresponding mass-to-light ratios do not differ notably from those characteristic of the oldest stellar populations that dominate this spectral range.

Figure 1 and Extended Data Figure 4 show the variation of the indices used for the fitting process as a function of the MLWA of a single SSP for two different metallicities: solar metallicity [M/H] = 0 (solid line), and super-solar metallicity [M/H] = +0.22 (dashed line). This illustrates the behaviour of each index with the population parameters. The observed indices are shown as black dots with 1-$\sigma$ level uncertainties. The bluer the index, the more it deviates from a single SSP model, which is a result of the additional young component that contributes more at shorter wavelengths. In contrast, the two reddest indices, H$\beta_o$ and [MgFe]' are quite insensitive to the younger component. The 2SSPs best-fitting indices are shown as red stars and show very good agreement with the observations. These fits require a combination of a small mass fraction in young stars (<1 Gyr) on top of an old component (see Figure 2 and Supplementary Table 2), which increases towards lower velocity dispersion. There is a strong anti-correlation between the derived mass fractions and ages of the young stars with velocity dispersion, with Spearman's correlation coefficient of –0.98 and –0.96, respectively. As a check, we performed the fits when changing the upper prior of the age of the old component as the MLWA to find that, despite the quality of the fits decrease significantly (> 2 times higher), the young mass fractions are lower than 0.1% in all mass bins and the trend with mass is less prominent. We also obtain lower young component ages, an effect of the burst age – burst strength degeneracy, which does not affect the 1SSP+cSFR model as described below.

Some of the indices fitted, such as Mg2852, NH3375 and Mg3334, shown in panels b) d) and e) of Extended Data Figure 4, respectively, are slightly affected by the overabundance of alpha-elements in ETGs, like Magnesium, or by the overabundance of Nitrogen[49]. However, the models used here are scaled-solar and there are no empirical stellar population models available in the literature with alpha-abundances covering the UV range. The deviation shown between the best-fitting solutions and the observed values increases with velocity dispersion. Such a result is expected from the well-studied correlation between Mg and σ in that Mg strongly increases with $\sigma^{39}$. When excluding these indices, the best-fit results do not change significantly, but the $\chi^2$ slightly decreases, as expected. Note that these indices are not perfectly fit because they also show a deviation with respect to a pure old SSP model. Although we cannot quantify properly the effect of varying the abundances, as no such models are currently available in the UV, the best fits show the expected mismatch. For example, Mg3334 would get stronger when increasing [Mg/Fe]. Note that the UV continuum flux of a stellar population increases with increasing alpha abundance. For massive ETGs with alpha-enhanced stellar populations, the young mass fractions would represent a lower contribution than the one found in this work, if alpha-enhanced models are used.

The best 2SSPs model spectrum for the highest velocity dispersion stack is illustrated in the panel a) of Extended Data Figure 5. The solution only requires 0.09%, in mass, of a young (~65 Myr) component, in addition to an old population, to match the stacked spectrum reasonable well, as indicated by the ratio in the panel b) of the figure. Note that this ratio does not represent residuals from a full spectrum fitting. For comparison, we also show the ratio between the 2SSPs and a pure old SSP model spectrum in the panel b). This ratio shows that the stack it is not well-resembled, especially to bluer spectral regions, by a pure old SSP and requires an additional young component. Note the significant impact of such a small fraction of the young component in the composite two SSPs spectrum at NUV wavelengths, which dominates the total flux budget at $\lambda \lesssim 3000$ Å. In the optical range, this young contribution has an insignificant effect and the main contribution comes from the dominant old stellar population. It is important to emphasize that we do not use all the information present in the spectrum, and therefore some residuals are to be expected. This is particularly true at those wavelengths that are strongly affected by, for example, abundance issues such as in the regions of the Mg2800 and Mg2852 features. Note also that the SNR decreases towards the bluest end of the spectrum which is reflected in the increasing residuals in this spectral region.

**Fitting approach 2: Stellar populations formed over the past 2 Gyr of galaxy evolution.** The 1SSP+cSFR approach assumes a composite population with constant SFR for the young component. The old component is described by a SSP, as in our first approach (i.e., using 2 parameters that cover the same range of age and metallicity). This is a valid approximation, as the expected age spread in the old component is comparatively narrower than the lookback time. Given the rapid formation of the bulk of the stars in massive ETGs, the residual star formation at late times is most likely produced by material ejected by evolved stars as well as available gas in the interstellar medium[27]. We tested different age intervals and found that limiting the duration of constant SFR between 6.3 Myr and 2 Gyr gives the best fit to the data. However, the results do not change significantly when varying the upper limit down to 1 Gyr (although the $\chi^2$ becomes slightly worse). The spectrum of the young component is obtained by integrating the individual SSPs with ages between 6.3 Myr and 2 Gyr, spaced logarithmically over time. This spectrum is added to an old stellar population, leaving the relative mass contribution as an additional free parameter (giving a model described by three parameters). The best-fit results for the young mass fractions are shown in the panel c) of Figure 2. We found that the contribution from the young stellar populations formed between 6.3 Myr and 2 Gyr is 0.48% at the highest velocity dispersion, weakly increasing towards the lower mass stacks, up to 0.58% for the lowest velocity dispersion stack. Note that although small, the increasing trend is significant in terms of light contribution to the NUV spectrum.

Results from the 1SSP+cSFR model do not experience significant variations when the prior is shifted to younger ages than 9.4 Gyr (i. e., setting the prior on the old component to the youngest possible age which is the MLWA). This is because only stellar populations with ages lower than 2 Gyr are those that significantly contribute to decrease the NUV line-strengths. Therefore, the mass fractions of stars formed in the last 2 Gyr are mostly independent of intermediate and old age stellar populations, which make these results very robust.

**Testing different SSP models.** We have also used the Bruzual and Charlot (B&C) models[22] to quantify the systematics associated with using different SSP models. Using the same methodology employed for E–MILES, the B&C's models give 2 Gyr younger MLWA. As a consequence, the subsequent mass fractions obtained with the 1SSP+cSFR model approach are slightly higher than those found with the E–MILES models, although the trend with mass remains: from 0.78% to 0.95% for the lowest mass stacks. However, the quality of the fits decreases considerably using the B&C's models, yielding a reduced $\chi^2$ of ~50.

**Other potential contributions.** In addition to young stars, other stellar components may contribute to the NUV spectra of massive ETGs, namely hot-evolved stars, such as those in the Post-Asymptotic Giant Branch (PAGB[14]. In this phase, the hot cores of stars are exposed and shine bright in the UV. As a test, we fit the data with models that combine an old stellar population with the spectrum of a high temperature (10,000 – 25,000 K) star, simulating a PAGB star. This yields a 4 – parameter model, which are the age and metallicity of the old stellar population plus the temperature and the relative V-band light contribution of the hot-old stellar component. On average, a ~3.8±0.07% light contribution in the V-band from a star with $T_{eff}$ ~13,600 K is required to fit the line-strengths, on top of an old component. Importantly, this light fraction is substantially higher than the values predicted by the stellar evolution theory, in the framework of stellar population synthesis[20,21,50], which predict significantly less than 1% of the flux in the V–band from the PAGB. Additionally, the $\chi^2$ values increase notably at lower masses for the PAGB model approach, with significantly higher values (by a factor of more than two at the lowest mass bin) with respect to the 2SSPs models. We explore the Bayesian Information Criterion[51] (BIC) model selection criterion to choose between both models. Except for the two most massive bins, one selects the young component model where the difference between both BIC values $\Delta BIC = BIC_{PAGB} - BIC_{2SSPs}$ is $\Delta BIC > 2$. Moreover, the obtained Teff~13,600 K is low, which corresponds to the first stages after envelope ejection. This is an extremely rapid phase, making it unlikely to contribute significantly. Hence, from all these arguments and selection criteria we conclude that even if a PAGB component were present, it has to be sub-dominant with respect to the very young stellar population we detect in the NUV. Our inferred young mass fractions, albeit upper limits, are not affected substantially by a PAGB-like stellar population component.

**Characterizing the environment of BOSS galaxies.** Abundance matching (AM) provides a simple way to statistically relate the observed galaxies to their dark matter halo hosts. AM models[52,53] compare the observed stellar mass function of galaxies with the halo mass distribution from N-body numerical simulations. We used the functional form provided by Behroozi[54], interpolating between their estimates at z=0.1 and z=0.5 to assess the expected correlation between stellar mass and halo mass at the median redshift of our sample (z~0.38). We created, via Monte Carlo sampling, a random set with the same number of galaxies as in our sample and with the observed stellar mass distribution of BOSS galaxies[29]. If we assume galaxy clusters inhabit halos more massive than $\sim 10^{14}$ M$_\odot$, we obtain a cluster fraction of ~1/2, i.e., approximately half of our BOSS galaxies live in clusters.

**Contrasting with numerical simulations of galaxy formation.** Our observational results are compared with state-of-the-art numerical, cosmological simulations. We employ the publicly available database from the EAGLE simulation project[55], that consists of catalogues of galaxies extracted from cosmological, hydrodynamic simulations. The EAGLE model of galaxy formation, includes a pressure law for star formation[56], line cooling in photoionisation equilibrium[57], stellar evolution[58], thermal supernova feedback[59] and massive black hole growth and feedback[60]. An extensive description of the model, its calibration and the hydrodynamic solver are given in several works[6,61,62].

The EAGLE reference simulation, used here, reproduces several global relations of observed galaxies at the current epoch and earlier redshifts. It consists of a cosmological cubic volume with 100 comoving Mpc on a side, and contains $2 \times 1504^3$ dark matter and baryon particles. The initial conditions have been evolved from z=127 to z=0. We use the snapshot at redshift z=0.365, the closest in time to the observed median redshift. The standard database provides most of the global properties of simulated galaxies, whereas we use the full particle data to compute the rest frame integrated spectra, line-of-sight velocity dispersions and SFHs of the synthetic galaxies.

The luminosities are computed following a method similar to that used by the EAGLE team[63], but employing the same E-MILES spectral library of the observational analysis. Fluxes from single SSPs (stellar particles in the simulation, characterised by their initial mass, metallicity and age) are derived by convolving the synthetic spectra with the response of the SDSS passbands (*u*, *g*, *r*, *i* and *z*). We avoid any extrapolation outside of the metallicity and age range of the population synthesis model by limiting the metallicities and ages of the simulation data. We include the effect of intrinsic dust attenuation following the standard modelling[63], but normalizing the extinction law with the surface density, inside the stellar half-mass radius, of heavy elements in the (star-forming) gas with hydrogen number density above ~0.1 cm$^{-3}$, instead of the gas total mass. This prescription produces reasonable attenuation values over a wide range of stellar masses and is compatible with the observational constraints of attenuation in massive ETGs. Following the EAGLE team methodology[63], recent (<100 Myr) star formation is resampled with finer time intervals, alleviating the effects of stochasticity of the star formation model. All spectra are computed within a three-dimensional aperture of 10 kpc, to better match the observed sample.

For a better comparison with the observed stacks, we compute the mass-weighted stellar velocity dispersion along the line of sight within a cylindrical aperture of 10 kpc. Finally, we extract the SFHs of each galaxy in the sample by binning the initial stellar mass of the SSPs as a function of their formation time in bins of 50 Myr and retrieve from the database the mass-weighted age of the galaxies.

**Code availability.** To measure line–strength indices of the spectra we use our own code based on the publicly available *indexf* code from https://indexf.readthedocs.io/en/latest/index.html. The *emcee* package used to fit the observed line-strength indices is available at https://emcee.readthedocs.io/en/stable/. Codes generated in this study are not publicly available since they can be reproduced following the Methods description but are available from the corresponding author upon reasonable request.

**Data availability.** The observational data used in this study is derived from the SDSS-III DR12 Survey Science Archive Server (https://dr12.sdss.org/advancedSearch). The stacked spectra data analysed during the current study that supports the plots within this paper are available as Source Data. The EAGLE simulation data is publicly available (http://eagle.strw.leidenuniv.nl/) and the simulated spectra of the EAGLE galaxies analysed in this manuscript are available from the authors upon reasonable request. The E-MILES SSP models are publicly available at the MILES website (http://miles.iac.es).

**Acknowledgements.** We thank Gustavo Bruzual and Stephano Charlot for kindly providing us their new version of stellar population models extended to the UV spectral range in order to test how results change when using other SSP models. Parts of this research have been funded by the Spanish Ministry of Science, Innovation and Universities (MCIU) through research project SEV-2015-0548-16-4 and predoctoral contract BES-2016-078409. N.S.R, M.A.B, A.V. and I.F. acknowledge support from grant AYA2016-77237-C3-1-P from the MCIU. FLB acknowledges financial support from the European Union's Horizon 2020 research and innovation programme under the Marie Sklodowska-Curie grant agreement n. 721463 to the SUNDIAL ITN network. M.A.B. and A.N. acknowledge the Severo Ochoa excellence scheme SEV-2015-0548. CDV acknowledges financial support from MCIU through grants AYA2014-58308 and RYC-2015-1807. We thank the anonymous referees for their valuable comments.

Funding for SDSS-III has been provided by the Alfred P. Sloan Foundation, the Participating Institutions, the National Science Foundation, and the U.S. Department of Energy Office of Science. The SDSS-III web site is http://www.sdss3.org/. SDSS-III is managed by the Astrophysical Research Consortium for the Participating Institutions of the SDSS-III Collaboration including the University of Arizona, the Brazilian Participation Group, Brookhaven National Laboratory, Carnegie Mellon University, University of Florida, the French Participation Group, the German Participation Group, Harvard University, the Instituto de Astrofísica de Canarias, the Michigan State/Notre Dame/JINA Participation Group, Johns Hopkins University, Lawrence Berkeley National Laboratory, Max Planck Institute for Astrophysics, Max Planck Institute for Extraterrestrial Physics, New Mexico State University, New York University, Ohio State University, Pennsylvania State University, University of Portsmouth, Princeton University, the Spanish Participation Group, University of Tokyo, University of Utah, Vanderbilt University, University of Virginia, University of Washington, and Yale University.

**Author contributions.** N.S.R lead the analysis and wrote the text. A.V., M.A.B. and F.L.B. developed the idea and supervised the work. I.F. prepared the BOSS data for analysis. F.L.B and N.S.R. wrote code. N.S.R, A.V., M.A.B, F.L.B. and I.F. interpreted and analysed the results. A.N. and C.D.V. prepared the simulation data. All authors discussed the results and edited on the manuscript.

**Correspondence.** Correspondence and request for materials should be addressed to N.S.R. (e-mail: nsalva@iac.es)

**Competing interests.** Authors declare no competing interests.

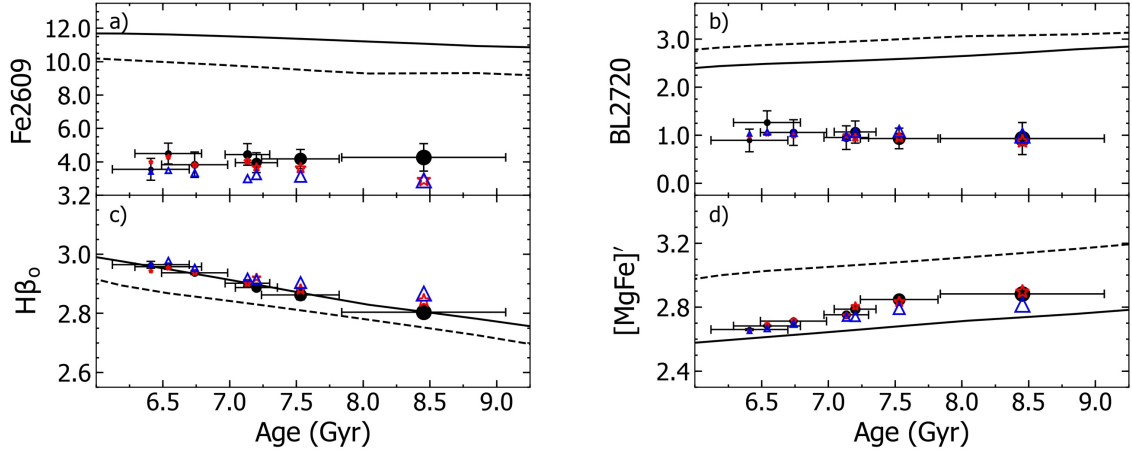

**Figure 1. NUV and optical line-strength indices.** The panels show the behaviour of selected NUV and optical spectral SSP model indices as a function of age, for two different metallicities ([M/H] = 0.0 (solar), solid lines; [M/H] = +0.22 (super-solar), dashed lines). Panels a) and b) show the NUV indices Fe2609 and BL2720 and panels c) and d) show the optical indices $H\beta_o$ and [MgFe]'. The black solid dots correspond to the observed measurements of BOSS stacked spectra of massive ETGs. Red stars indicate the best-fitting 2SSPs model predictions, that assume a superposition of an old and a young component, and blue triangles show the best-fitting 1SSP+cSFR model predictions, that models the young component as a composite population with constant SFR between 6.3 Myr and 2 Gyr. Error bars on the black dots indicate the 1-σ uncertainty on the index measurements. The impact of very small fractions of young stars on the NUV indices, with respect to a single SSP model, is significant. In contrast, the optical indices are unaffected by these residual populations. The SSP models and data are convolved to a common velocity dispersion of σ = 340 kms$^{-1}$. See Extended Data Figure 4 for a description of the other line-strength indices used in this work.

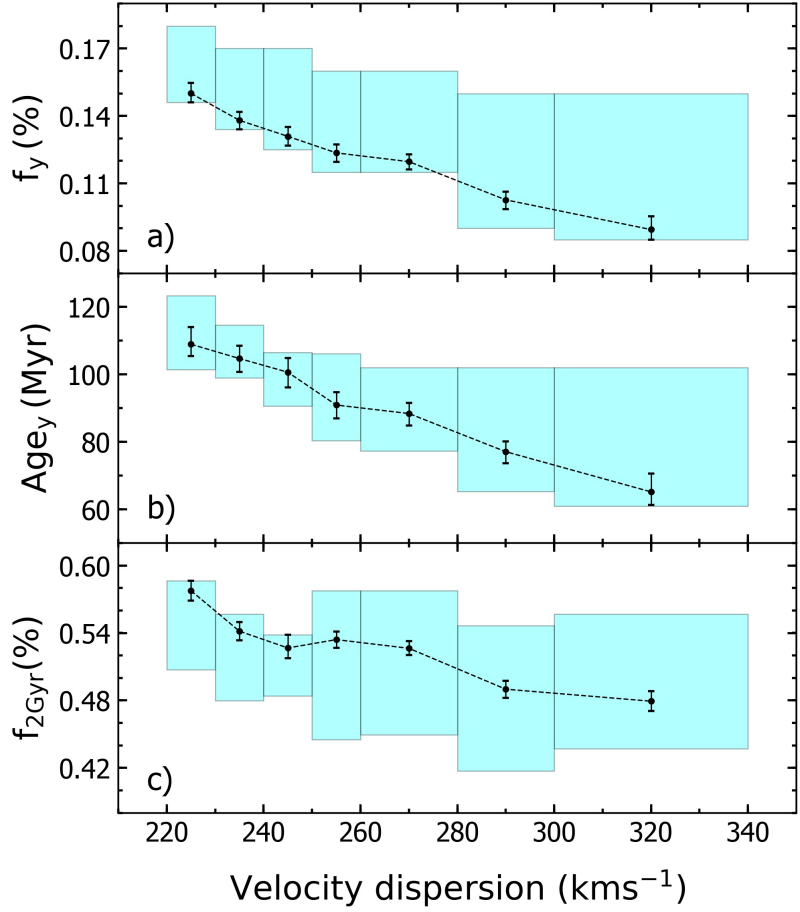

**Figure 2. Constraining recent star formation in massive ETGs.** The derived contributions from young stellar populations are shown as a function of the velocity dispersion range of the stacked BOSS galaxies. Black dots indicate the results from MCMC and error bars indicate 1-σ uncertainties obtained from marginalizing over the rest of the parameters. Panels a) and b) show the mass fraction in young stars and the mean luminosity-weighted age, for the 2SSPs model. Panel c) gives the mass fraction in young stars following the 1SSP+cSFR fitting approach. Note the decreasing young mass fraction with increasing velocity dispersion (or galaxy mass). Blue areas show the range of solutions when varying the weights of the optical and NUV set of indices from 100% to 0% in the $\chi^2$ minimization with the index uncertainties at the 1-σ level (see Methods).

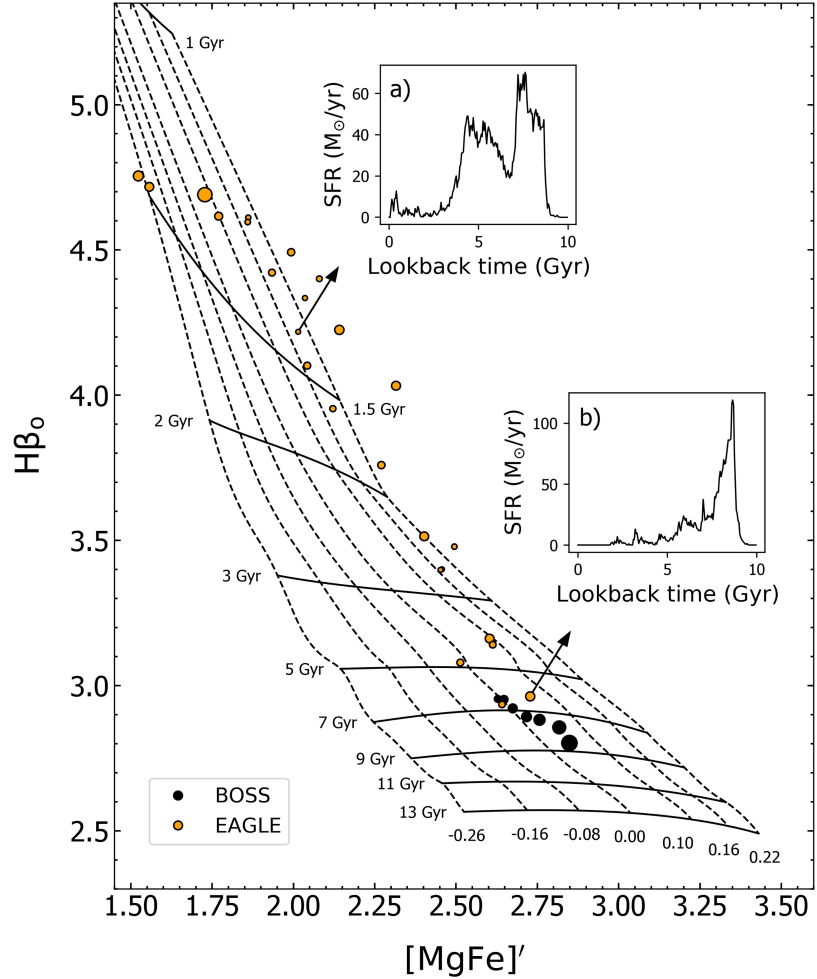

**Figure 3. Comparison with numerical simulations of galaxy formation.** Measurements of the age-sensitive spectral index $H\beta_o$ are plotted as a function of the total metallicity proxy [MgFe]'. The grid shows the model predictions over a range of SSP ages (solid lines) and metallicities (dashed lines), as labelled. The black dots are the observed BOSS index measurements, with dot size increasing with galaxy mass and uncertainties smaller than the point size. The orange dots represent synthetic galaxies from the EAGLE simulation, chosen to match the same selection criteria as in our BOSS data, except for the colour cut (g-i)>2.35, that only is satisfied by the two simulated galaxies closest to the BOSS data indices. Their symbol size increases with galaxy mass. The figure also shows the SFHs of a young and old simulated galaxy, in panels a) and b) respectively, illustrating why most of the simulated galaxies have younger ages than expected from the BOSS data. The model grid and both simulated and observational spectra are smoothed to a common velocity dispersion of 340 kms$^{-1}$.

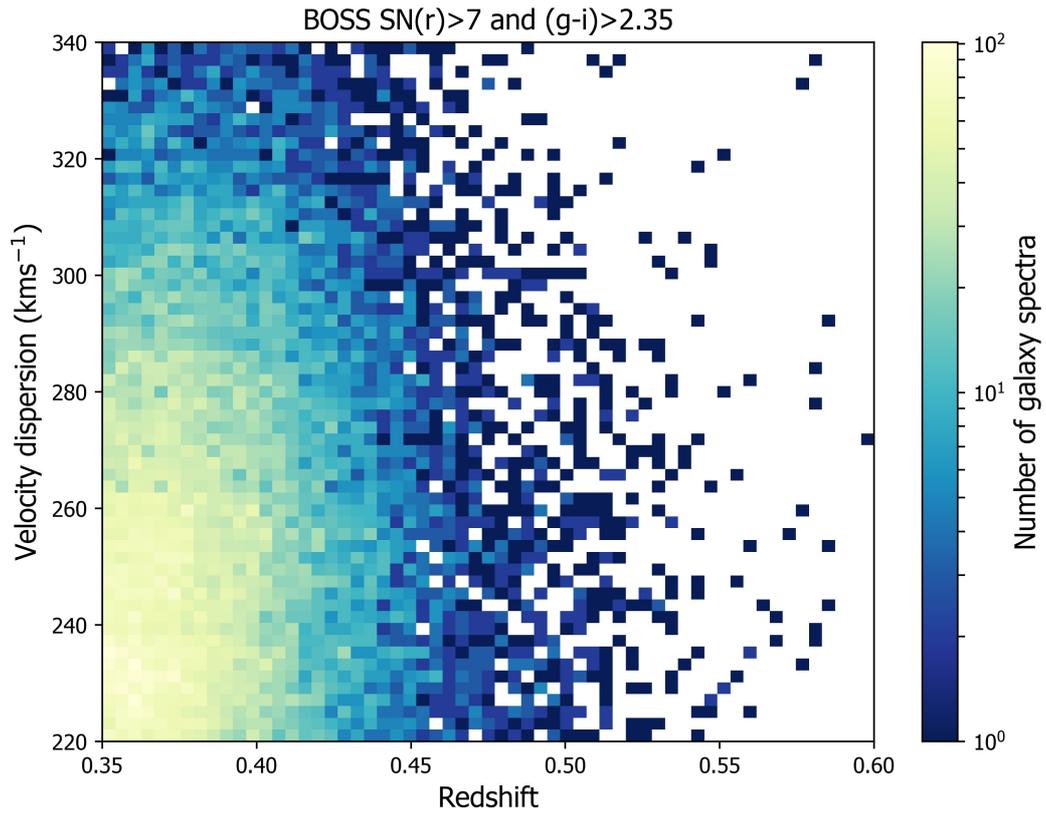

**Extended Data Figure 1. Data selection criteria.** Selection of our BOSS galaxy sample as a function of velocity dispersion and redshift. The plot illustrates the number density in logarithmic scale of available BOSS spectra with SNR>7 in the SDSS-r band and (g-i)>2.35. We select BOSS galaxies with redshift 0.35≤z≤0.6, and velocity dispersion between 220 and 340 kms$^{-1}$. Lowest density regions show individual galaxies with dark blue.

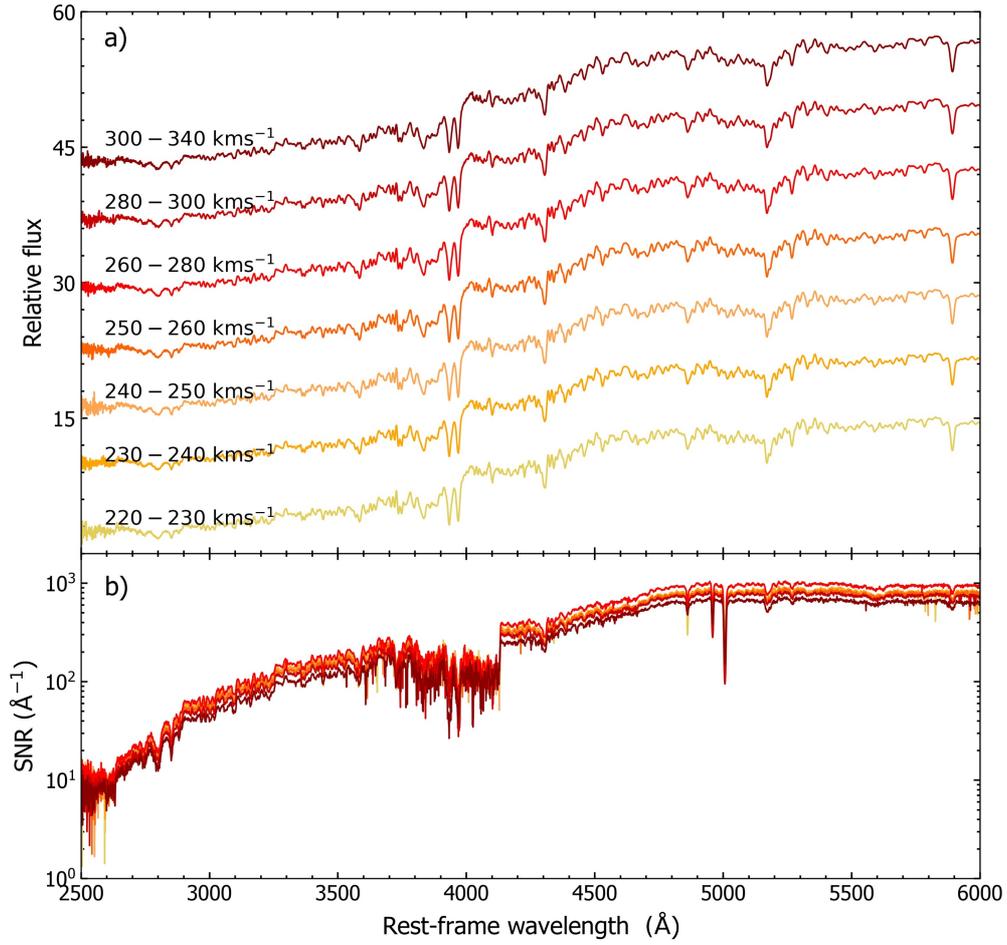

**Extended Data Figure 2. BOSS stacked spectra.** Panel a) shows our stacked spectra from the BOSS survey, colour-coded with respect to the stellar velocity dispersion range labelled on the top-left of each spectrum. The spectra have been shifted vertically for ease of visualization. Panel b) represents the signal-to-noise ratio, following the same colour scheme.

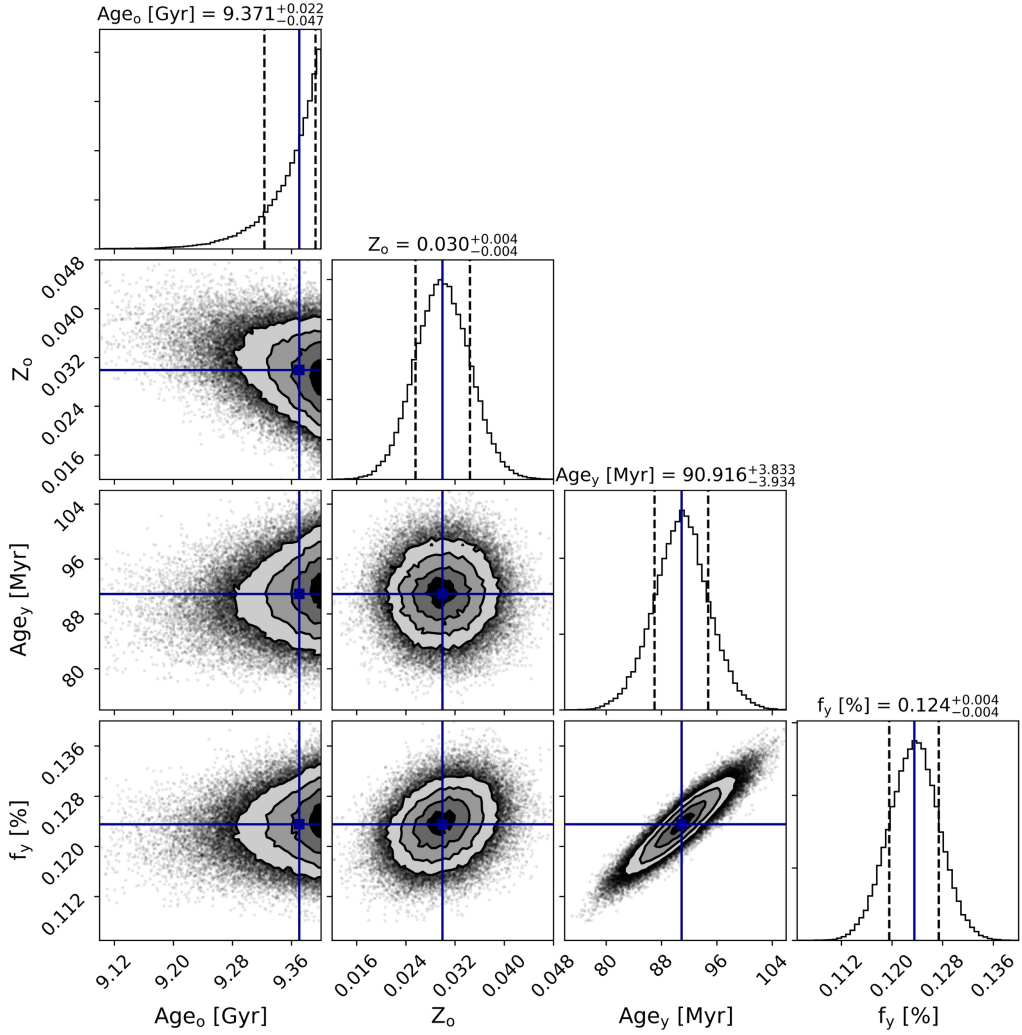

**Extended Data Figure 3. Probability distribution functions from MCMC.** Probability distribution functions of an intermediate velocity dispersion stack (250 – 260 kms$^{-1}$) corresponding to the model parameters obtained with the 2SSPs approach. From top to bottom: age and metallicity for the old component, and age and mass fraction of the young component. The panels show the marginalized PDF over each single parameter over the other parameters. Contours show the two-dimensional 1– σ, 2– σ and 3– σ confidence levels between all the parameters. The last panel of each row shows the median (blue solid line) and the 16 and 84 percentiles (dashed lines) for each PDF. Note the burst age – burst strength degeneracy found between the age and the fraction of the young component, for which a smaller burst can be produced more recently.

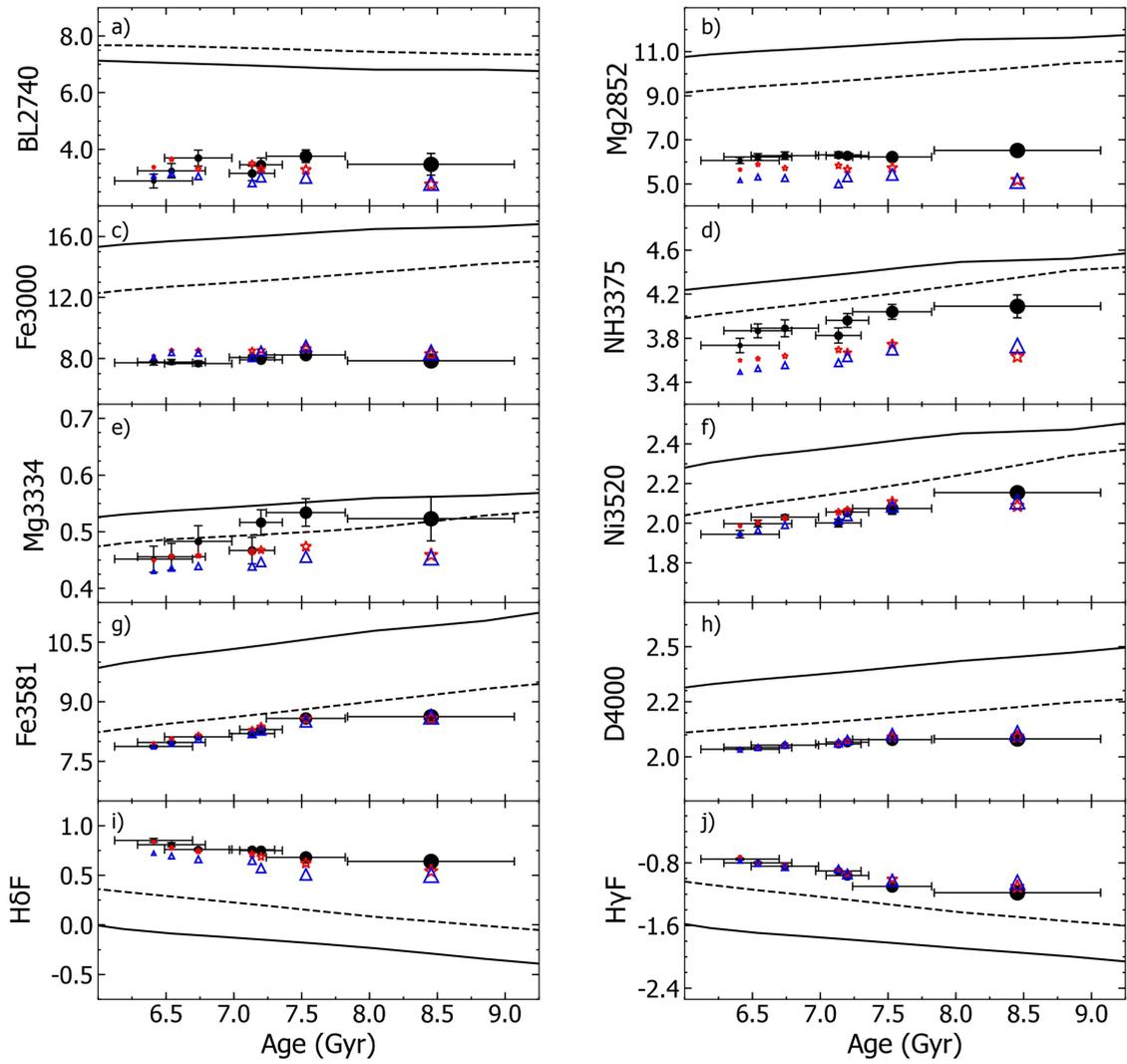

**Extended Data Figure 4. NUV and optical line-strength indices.** NUV and optical line-strength indices. The panels show the behaviour of the NUV and optical SSP model indices as a function of age, for two different metallicities ([M/H] = 0.0, solid lines; [M/H] = +0.22, dashed lines). The black dots correspond to the observed measurements of BOSS stacked spectra and their error bars show the 1−σ uncertainty on the measurements. Red stars indicate the best-fitting 2SSPs model predictions. Blue triangles show the index values predicted by the best-fitting 1SSP+cSFR model. Note that the discrepancy between SSP model predictions and data increases towards bluer indices. The indices shown exclude those in Figure 1.

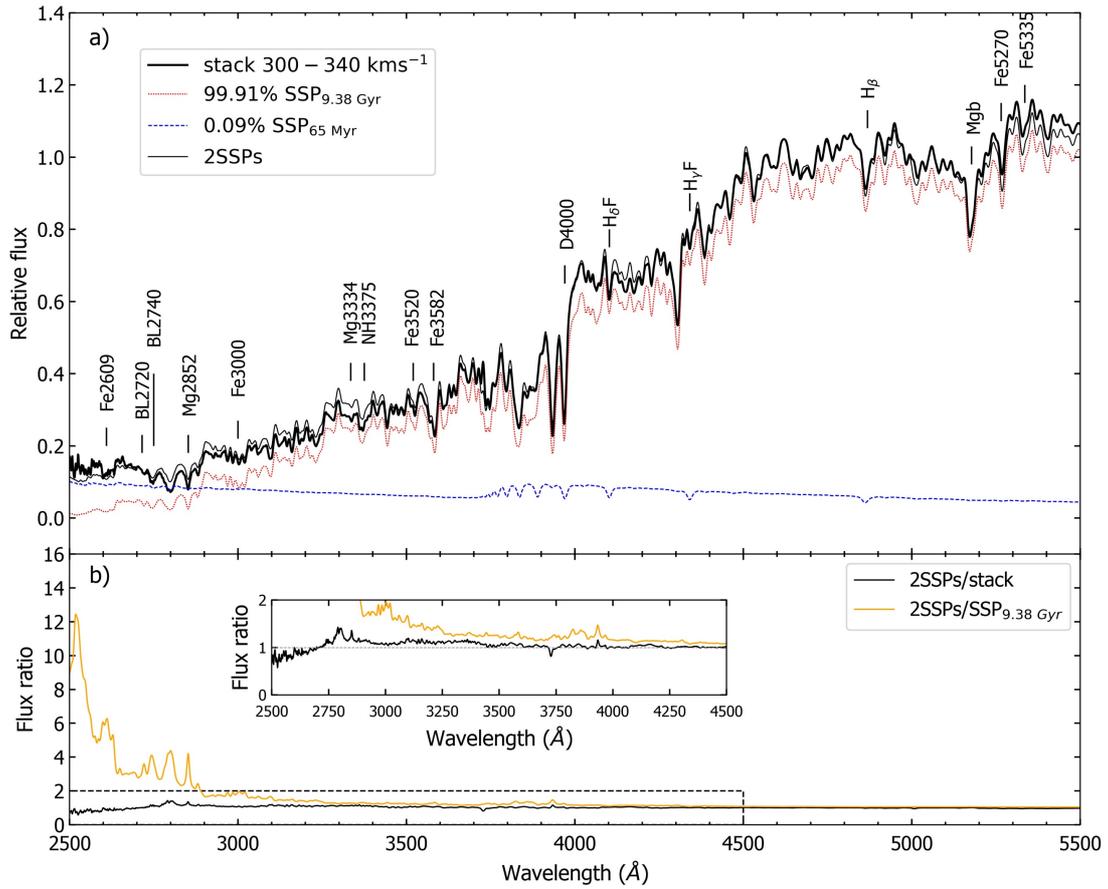

**Extended Data Figure 5. 2SSPs best index fit spectrum.** Panel a) shows the stacked spectrum with the highest velocity dispersion range (thick black line). Overplotted is the 2SSPs model (thin black) that best fits the 14 indices shown in Figure 1 and in Extended Data 4. The main purpose of this figure is to illustrate the relative effect of the young component as a function of wavelength but note it does not represent a full-spectrum fit. The 2SSPs model can be split into the young (65 Myr) component (blue dashed) which contributes a 0.09% mass fraction, and the old (9.38 Gyr) dominant population (red dotted). Panel b) shows the ratio between the 2SSPs best-fit model with the stacked spectrum (black) and with a pure old SSP model of 9.38 Gyr (orange). The inset subplot shows the flux ratio of the dashed box in more detail.

| Index | Blue Passband | Index Passband | Red Passband | Reference |
|---|---|---|---|---|
| Fe2609 | 2,562.000-2,588.000 | 2,596.000-2,622.000 | 2,647.000-2,673.000 | 1,2,3 |
| BL2720 | 2,647.000-2,673.000 | 2,713.000-2,733.000 | 2,762.000-2,782.000 | 1,3 |
| BL2740 | 2,647.000-2,673.000 | 2,736.000-2,762.000 | 2,762.000-2,782.000 | 1,3 |
| Mg2852 | 2,818.000-2,838.000 | 2,839.000-2,865.000 | 2,906.000-2,936.000 | 1,2,3 |
| Fe3000 | 2,906.000-2,936.000 | 2,965.000-3,025.000 | 3,031.000-3,051.000 | 1,2,3 |
| NH3375 | 3,342.000-3,352.000 | 3,350.000-3,400.000 | 3,415.000-3,435.000 | 4 |
| Mg3334 | 3,310.000-3,320.000 | 3,328.000-3,340.000 | 3,342.000-3,355.000 | 4 |
| Ni3520 | 3,499.000-3,508.000 | 3,511.000-3,530.000 | 3,532.000-3,548.000 | 1 |
| Fe3581 | 3,532.000-3,548.000 | 3,548.000-3,594.000 | 3,594.000-3,602.000 | 5 |
| D4000 | 3,750.000-3,950.000 | 4,050.000-4,250.000 | | 6 |
| H$\delta$F | 4,057.250-4,088.500 | 4,091.000-4,112.250 | 4,114.750-4,137.250 | 7 |
| H$\gamma$F | 4,283.500-4,319.750 | 4,331.250-4,352.250 | 4,354.750-4,384.750 | 7 |
| H$\beta_o$ | 4,821.175-4,838.404 | 4,839.275-4,877.097 | 4,897.445-4,915.845 | 8 |
| Mgb5177 | 5,142.625-5,161.375 | 5,160.125-5,192.625 | 5,191.375-5,206.375 | 9 |
| Fe5270 | 5,233.150-5,248.150 | 5,245.650-5,285.650 | 5,285.650-5,318.150 | 9 |
| Fe5335 | 5,304.625-5,315.875 | 5,312.125-5,352.125 | 5,353.375-5,363.375 | 9 |

**Supplementary Table 1. Definitions of the NUV and optical line-strength indices.** Each index is defined by a central passband and a blue and a red pseudo-continuum spectral ranges. The passband wavelengths are given in Angstroms. The combined optical index [MgFe]' is obtained via: [MgFe]' = $\sqrt{\text{Mgb5177}(0.72\text{Fe5270} + 0.28\text{Fe5335})}$ The references in the last column are 1: Fanelli et al. (1990)[64]; 2: Maraston et al. (2009)[65]; 3: Chavez et al. (2007)[66]; 4: Serven et al. (2011)[49]; 5: Gregg (1994)[67]; 6: Bruzual (1983)[68]; 7: Worthey & Ottaviani (1997)[69]; 8: Cervantes & Vazdekis (2009)[70]; 9: Worthey et al. (1994)[71].

| σ (kms$^{-1}$) | N | z | log(mass) | Age (Gyr) | [M/H] | Age$_o$ (Gyr) | [M/H]$_o$ | Age$_y$ (Myr) | f$_y$ (%) | f$_{2Gyr}$ (%) |
|---|---|---|---|---|---|---|---|---|---|---|
| 220-230 | 4180 | 0.376±0.030 | 11.32±0.16 | 6.6±0.2 | -0.008±0.009 | $9.35^{+0.04}_{-0.08}$ | $-0.013^{+0.004}_{-0.004}$ | $109^{+5}_{-4}$ | $0.150^{+0.005}_{-0.004}$ | $0.58^{+0.01}_{-0.01}$ |
| 230-240 | 4517 | 0.376±0.030 | 11.33±0.16 | 6.8±0.2 | -0.003±0.008 | $9.37^{+0.02}_{-0.05}$ | $-0.010^{+0.004}_{-0.004}$ | $105^{+4}_{-4}$ | $0.138^{+0.005}_{-0.004}$ | $0.54^{+0.01}_{-0.01}$ |
| 240-250 | 4136 | 0.378±0.031 | 11.35±0.16 | 7.2±0.4 | -0.005±0.015 | $9.38^{+0.02}_{-0.03}$ | $-0.004^{+0.004}_{-0.004}$ | $101^{+4}_{-4}$ | $0.131^{+0.005}_{-0.004}$ | $0.53^{+0.01}_{-0.01}$ |
| 250-260 | 4077 | 0.378±0.030 | 11.36±0.17 | 7.3±0.1 | 0.014±0.004 | $9.37^{+0.02}_{-0.05}$ | $0.030^{+0.004}_{-0.004}$ | $91^{+4}_{-4}$ | $0.124^{+0.005}_{-0.004}$ | $0.53^{+0.01}_{-0.01}$ |
| 260-280 | 5623 | 0.381±0.030 | 11.39±0.18 | 7.2±0.2 | 0.03±0.01 | $9.38^{+0.01}_{-0.02}$ | $0.052^{+0.004}_{-0.004}$ | $88^{+3}_{-4}$ | $0.119^{+0.005}_{-0.004}$ | $0.53^{+0.01}_{-0.01}$ |
| 280-300 | 3516 | 0.383±0.031 | 11.43±0.19 | 7.8±0.3 | 0.05±0.01 | $9.38^{+0.02}_{-0.03}$ | $0.082^{+0.005}_{-0.005}$ | $77^{+3}_{-3}$ | $0.103^{+0.005}_{-0.004}$ | $0.49^{+0.01}_{-0.01}$ |
| 300-340 | 2614 | 0.388±0.032 | 11.47±0.20 | 8.6±0.5 | 0.05±0.002 | $9.38^{+0.02}_{-0.04}$ | $0.095^{+0.006}_{-0.006}$ | $65^{+5}_{-4}$ | $0.089^{+0.005}_{-0.004}$ | $0.48^{+0.01}_{-0.01}$ |

**Supplementary Table 2. General properties and best-fitting parameter results of the BOSS stacked spectra.** They are shown with respect to the velocity dispersion range of the stacks (Col. 1). Col. 2 shows the number of individual galaxy spectra added in each bin. Col. 3 and Col. 4 show the average redshift and stellar mass calculated from spectral energy distribution fitting[29]. Col. 5 and 6 is the MLWA and metallicity derived from the H$\beta_o$ and [MgFe]' indicators. Col. 7 and 8 show the SSP age and metallicity of the old component. Col. 9 and 10 show the stellar mass fraction and age of the young component of the 2SSPs model. Col. 11 is the young mass fraction following the 1SSP+cSFR approach, with a constant SFR to describe the young component.